\documentclass[aps,prl,reprint,amsmath,amssymb,showkeys]{revtex4-1}
\usepackage{graphicx}
\usepackage{float}
\usepackage{dcolumn}
\usepackage{bm}%
\usepackage{enumerate}
\usepackage{tabularx}
\usepackage{subfig}
\usepackage{graphfig}

\begin{document}

\title{The upper bound of packet transmission-capacity in local static routing}


\author{Zhe He}
\altaffiliation{School of Physics, University of Science and Technology of China,Hefei 230026, People's Republic of China}
\email{2287580716@qq.com, ychezhe@mail.ustc.edu.cn}
\author{ Rui-Jie Xu}
\author{Si-Wu Liu}
\author{Bing-Hong Wang}
\email{bhwang@ustc.edu.cn}
\thanks{Thank professor Xinmao Wang, Jianliang Zhai and Benjin Xuan from University of Science and Technology of China. Thank professor Wenxu Wang from Beijing Normal University. Thank my fellows Yiming Huang, Benchuan Lin. This work is supposed by The National Natural Science Foundation of China.
The National Natural Science Foundation of China (Grant Nos. : 11275186, 91024026, FOM2014OF001 )}


\date{\today}

\begin{abstract}
We propose a universal analysis for static routings on networks and describe the congestion characteristics by the theory. The relation between average transmission time and transmission capacity is described by inequality $T_0Rc_0\leq1$. For large scale sparse networks, the non-trivial upper bond of transmission capacity $Rc_0$ is limited by $Rc_0\leq1/<1/k>$ in some approximate conditions. the theoretical results agree with simulations on BA Networks.
\end{abstract}


\maketitle

\section{\label{1}Introduction}
\subsection{\label{1.1}Overview}
With the development of information age, the increasing of network scale and the expanding of network function, studies of network structure, performance and dynamics have gained more and more attention.As network research springs up, a variety of routings on complex networks have been proposed\cite{YanGang,WangWenXu,WangWenXu2,LingXiang,LingXiang2}.They can be divided into four categories:
\begin{enumerate}
  \item local static routing, based on local structure information of network (e.g. degree),open-loop system(non-fedback)
  \item local dynamic routing, based on local state information (e.g.queue length), close-loop (Feedback)
  \item global static routing, based on entire (or a significant portion of) network structure information, open-loop system(non-fedback)
  \item global dynamic routing, based on state information of entire (or a significant portion of) network,close-loop (feedback)
\end{enumerate}
In terms of packets transmission performance, global dynamic routings are the best, for the routings are derived from the entire structure of network with real-time feedback, which equals to give us a lot of extra information. However, with the increase of the scale of various communication networks in reality and with the appearance of some special networks (e.g. wireless Self-organizing network\cite{IEEE,ACM},it is impossible for people to know entire or almost entire structure of a network, so that the local routings are especially important. We need a great amount of real time communications if we use a close-loop routing which may increase the burden of network transmission. Therefore, there are special values in finding a high-performance open-loop routing, especially in some networks with limited communication abilities.
\\
In the face of various routings proposed, there is a question: Given a certain network, respectively in four kinds of routings, how good can the optimal routing of each kind be? How can we find them?
\\
In this paper, we study the congestion characteristics of a routing on a network. We use transmission capacity of the data packets ($Rc$) to measure it and analyze the routings using Markov process. we present the upper bond of $Rc$ of all static routings and estimate the divergence speed of it as network scale N increases.
\subsection{\label{1.2}Models and routing rules}
In a communication network, information (data) is transmitted in packets. Each device with the ability of data packets generating, delivering, receiving and analyzing abilities (routers in Computer communication network, agents in wireless communication networks) is abstracted into a vertex in this model. The physical connections between them correspond to edges between the vertexes. Consequently, the networks in the model reflect the topological properties of real networks. The same model is used here as in\cite{YanGang,WangWenXu,WangWenXu2,LingXiang,LingXiang2}.
\\
The packet transmission rules in the model are as follow:
\\

Each time step, we select $C$(If less then $C$, then all) data packets from the head of the queue of each vertex's cache and transfer them to their neighbours (vertexes with edges connecting each other). Meanwhile, $R$ new data packets are generated in the network. Each packet is born randomly at any vertex and randomly selects a vertex as the destination (except the vertex itself).The cache of each vertex is a queue structure and the new coming packets are placed at the end of the queue (FIFO). In addition, if a packet comes from any vertex $i$, then the next $n$ steps, the packet does not return to $i$, we call it $n-avoiding$. When $n$ is small, especially for the routing on sparse networks with large $Rc$, the effect of $n-avoiding$ is always a small quantity. Thus, in this paper, we set $n=0$ in analysis and $n=1$ in simulations on BA networks\cite{Albert}.
\\
The arrival principal is as follow:
 \\
 if a packet arrives at the neighbour of its target vertex(destination), it will be transferred to the target vertex at next step. if a packet arrives at its target vertex, it will be removed from the network.
\\
The static routing is as follow:
 \\
 for any vertex $i$, at any step $t$, we transfer the data packets at the head of the queue of $i$¡¯s cache to the vertex's neighbours vertex $j$ ($j=1,2,3¡­$) with probability $p_{ij}$, when the target vertexes of these packets are not vertex $i$'s neighbours.Transition matrix $P=(p_{ij})_{n\times n}$ is given by local information of network structure. If $p_{ij}$ is related to $t$ and $P(t)=(p_{ij}(t))_{n\times n}$ (usually $P$ is not explicitly a function of $t$ and relies on $t$ through system states), we call it local dynamic routing. If $p_{ij}$ is related to target vertex $x$ and $P=(p_{ijx})_{n\times n\times n}$ is given by global information of network structure, we call it global static routing. If $p_{ij}$ is related to $t$ and $x$, and $P=(p_{ijx}(t))_{n\times n\times n}$, we call it global dynamic routing.
\\
The definition of various routings in this paper is a widespread frame type definition. Almost all the routing strategies studied till now can be included in our frame. Such definition shows significance to comprehend the essence of routing problems.
\\
The meaning of transition matrix $P$ here is: bandwidth. A bandwidth refers to the maximum amount of information transferring through a channel per unit time. There is a little difference between this definition and what we discuss here. In this paper, the bandwidth means average bandwidth, which is the average amount of information transferring through a channel per unit time.Here, bandwidth is not only determined by the physical properties of a channel. We can properly design the allocation of the bandwidth of the whole network, which means we can design the amount of information transferring through a channel per unit time manually.
\\
In fact, the problem, finding optimal routing, is a process of optimization of network bandwidth allocation. For local routing strategies, the method is obvious. For global routing strategies, we can solve it form different methods such as finding the optimal path by different meanings and generating the corresponding routing table. At the time when the routing table is generated and the packet generation rate of each vertex is given, a bandwidth allocation is already determined. Through different routing algorithm, we can get difference routing strategies and naturally generate different allocations of the bandwidth.
\\
Model assumptions: 1.Network structure is constant or slowly varying. 2.The cache of the device is large enough that \textquoteleft out of memory\textquoteright ~will not happen. 3.Packets born time is stable. 4.Data packets are uniformly born in each vertex, and equiprobably choose their target vertexes.
\\
When $R$ is relatively small, as time goes on, the total number of data packets in network (the number of packets that has not yet arrived at the target vertexes at the current time) will reach a constant. This constant is of course limited, which has been supported by the results of computer simulations. The dynamic progress that can reach a stationary state like this is called free flow phase. However, when $R$ is relatively large, precisely, $R$ is larger than some $Rc$, the number of packets in network will increase with time and tend to infinity. Such dynamic process is called congestion flow phase. We measure network transmission capacity ($Rc$) by whether there is congestion or not. Quantitatively, $Rc$ is defined as the transition point of order parameter\cite{9}
\[\eta  = \mathop {\lim }\limits_{t \to \infty } \frac{{C\Delta W}}{{R\Delta t}}\]
where $\Delta W = W(t + \Delta T) - W(t)$~expressed the number of data packets in network at t time step\cite{Arenas}.
\\
Our mission is under limited information, making $Rc$ as large as possible. In fact, $Rc$ always affects another quantity--average transmission time $T$. Precisely, we should find maximum $Rc$ with limited information and given $T$.
\\
\section{\label{2}Model analysis}
\subsection{\label{2.1}Analytical method and some results}
First we have the follow conclusion: When the system has reached its stationary state, given a packet whose target(or birth vertex) is a certain vertex, the probability of finding it in any position of a queue of a given vertex is equal. Because when the system reaches a stationary state, it is a non time varying stochastic system.
\\
Let $\alpha_{ij}$ means after the system reaches the stationary state, at the head of the queue of vertex $i$, the probability of finding a packet whose target is vertex $j$. We use the relation that in average, the packet numbers of arriving and leaving are equal for stationary state, then we get the following equation:
\[{\alpha _{ij}} = \sum\limits_k {{p_{ki}}{\alpha _{kj}}(} 1 - {a_{kj}}) + \frac{R}{{CN(N - 1)}},i \ne j\]
\[{\alpha _{ii}} = 0\]
So
\begin{equation}
\alpha  = {P^T}\alpha  - {P^T}(A \circ \alpha ) + \frac{{RJ}}{C}
\label{eq:0}
\end{equation}
Where $¦Á$, $P$, $J$, $A$ refer to square matrices of order $N$. $P$ is a transition matrix. $P^T$ refers to the transpose of $P$. $A$ refers to an adjacency matrix of network. $\circ$ means Hadamard product. Each element of the Hadamard product of two matrices is the product of the corresponding elements of the two matrices. $C\alpha$ suggests the leaving packets each step. $P^T\alpha-P^T(A\circ\alpha)$ suggests the incoming packets, where $-P^T(A\circ\alpha)$ suggests that if a packet arrives at its target's neighbours, it will be directly transferred to its target.
\\
$J$ is a born matrix. $RJ_{ij}$ is expectation of the number of packets born in vertex $i$ whose target vertexes are vertex $j$ per time step. In this paper, we suppose that the packets born at different vertexes whose targets are not their own born vertexes are distributed with equal probability.
So, we have:
\[J = \frac{1}{{(N - 1)N}}\left( {\begin{array}{*{20}{c}}
0&1& \cdots &1\\
1& \ddots & \ddots & \vdots \\
 \vdots & \ddots & \ddots &1\\
1& \cdots &1&0
\end{array}} \right)\]
With both sides divided by R/C, Eq.(\ref{eq:0}) can be rewritten as:
\begin{equation}\label{eq:1}
{\alpha _0} = {P^T}{\alpha _0} - {P^T}(A \circ {\alpha _0}) + J
\end{equation}
where ${\alpha _0} = \frac{{C\alpha }}{R}$.
We suppose:
\begin{equation}\label{eq:2}
P{d_i} = (I - diag({a_{i1}},{a_{i2}} \cdots {a_{iN}}))P
\end{equation}
where $I$ refers to $N$ order unit matrix. This operation sets the corresponding rows of vertex $i$ and its neighbours to zero in matrix $P$. The physic meanings are obvious: When a packet whose target is vertex $i$ arrives at vertex $i$ or its neighbours, it will not be transferred to any vertexes expect vertex $i$ and will be immediately removed from network (After arriving the target vertex).
let column $i$ of matrix $\alpha_{0}$ be $\beta_{0i}$, column $i$ of matrix $J$ be $Ji$:
\[{\beta _{0i}} = P{d_i}^T{\beta _{0i}} + {J_i}\]
that is:
\begin{eqnarray}\label{eq:3}
\begin{array}{c}
{\beta _{0i}} = {(I - Pd_i^T)^{ - 1}}{J_i} = (I + Pd_i^T + {(Pd_i^T)^2}  +  \cdots ){J_{i}}\\
i = 1,2,3 \ldots N
\\
\\
\end{array}
\end{eqnarray}
For any $i$, if and only if the series of Eq.(\ref{eq:3}) converges, Eq.(\ref{eq:1}) has a unique solution.
\\
A sufficient condition of the convergence is : the graphs with adjacency matrix $A$ and $P$ are strongly connected. This is a natural requirement the problem we study must satisfy.
\\
If and only if $a_{ij}\neq0$, $p_{ij}\neq0$. Then we say non-negative matrix $P$ is consistent with $A$.
\\
In this paper, we only discuss the situation that Eq.(\ref{eq:1}) has a unique solution.
\\
From Eq(\ref{eq:3}), it¡¯s easy to know the solution of the Eq.(\ref{eq:1}): all the elements of matrix $\alpha$are non-negative.
\\
Let row $i$ of $\alpha$ be $\alpha_{i}$, then $s_{i}=\alpha_{i}\times1$ presents the probability of the existence of packets at the head of queue of vertex $i$ after the process reaching a stationary state.
So, the requirement that there is a stationary state is: $max\{s_{i}\}\leq1$,~the critical state satisfies $max\{s_{i}\}=1$,~equivalently, $||\alpha||_{\infty}=1$.~Here $Rc=C/||\alpha_{0}||_{\infty}$.
\\
Now, we have analyzed network transmission capacity $Rc$.  Given an adjacency matrix presenting network structure information and a transition matrix $P$ presenting routing strategy, we can calculate $Rc$ using Eq.(\ref{eq:1}).
\\
Let the average transmission time of packets be $T$, we have:
\[\begin{array}{l}
T = \mathop {{1^T}\sum }\limits_u \mathop \sum \limits_v [u \times ({(Pd_v^T)^{u - 1}} - {(Pd_v^T)^u}){J_v}]1\\
 = \mathop \sum \limits_{ij} {\alpha _0} = {1^T}{\alpha _0}1
\end{array}\]
where $1^T$ refers to ${(\begin{array}{*{20}{c}}
{\begin{array}{*{20}{c}}
1&1&1
\end{array}}& \cdots &1
\end{array})}$
\\
Accordingly, the average transmission time of packets whose targets are vertex $s$ is:
\[{T_s} = N \times {1^T}\mathop {\sum [}\limits_u u \times ({(Pd_s^T)^{u - 1}} - {(Pd_s^T)^u})]{J_s} = N{1^T}{\beta _{0s}}\]
\\
Now, we surprisingly find the significance of matrix $\alpha_{0}$. The reciprocal of the sum of its maximum line reflects the network transmission capacity. While the sum of each row multiplying the vertex number $N$ equals to the average transmission time of the packets whose target is the corresponding vertex. Transmission capacity and average time are two of the most important indices of transmission properties of network. Matrix $\alpha_{0}$  can describe these two clearly and can also quantitatively present the relation between them.
\\
According to the meaning of $\alpha_{ij}$ and the uniformity assumption above, approximately we have: \[{L_i} = C \times {s_i}\]
where $L_{i}$ represents the average queue length of vertex $i$ after reaching the stationary state.
\\
The whole length~\[L = \mathop \Sigma \limits_i {L_i}\]
So
\begin{equation}\label{eq:4}
\frac{L}{R} = \mathop \sum \limits_{ij} {\alpha _{0ij}} = T
\end{equation}
These represent the relation among the transmission time of packets (waiting time), number of data packets born per time and length of queue of each vertex. This relation agrees with the classical conclusion\cite{ross,Stewart}.
\begin{figure}[!ht]\centering
  \includegraphics[scale=0.3]{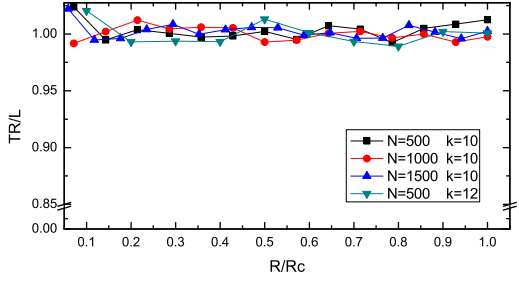}
  \caption{(color online) The abscissa refers to $R$ normalized by $Rc$. $T$ refers to average transmission time and $L$ refers to the sum of time average queue lengths of all vertexes. Each data point is the average of simulation results on $10$ networks when system reaches its steady state.}\label{fg:1}
\end{figure}
\\
When the network size $N$ is large, in Eq.(\ref{eq:0}), the elements of matrix $J$ tend to zero at the speed of $1/N$ and the elements of $\alpha$ are at the magnitude of $1/N$. The number of subtrahends in Eq.(\ref{eq:2}) proportion of the total element also tends to zero at the speed of $1/N$ and $R$ is remarkably less than $Rc$. Eq.(\ref{eq:0}) asymptotically becomes~ $\alpha1=P^T\alpha1$~with $N$ increasing.
\\
The solution of $\alpha1=P^T\alpha1$~is the stationary distribution of Markov process $P$ (differing from a positive scale coefficient). We sign the stationary distribution as $\pi=(\pi_1,\pi_2,\pi_3 ,\cdots\pi_n)$. Then $L_i\propto\pi_i$
Rewrite Eq.(\ref{eq:4}) into
\begin{equation}\label{eq:5}
T = \frac{C}{{{\pi _m}{Rc}}}
\end{equation}
Where $\pi_m$~refers to the maximum component of stationary distribution $\pi$.
\\
In order to see the essence clearly, we let \[{Rc_{0}} = \frac{{{Rc}}}{C}, {T_0} = \frac{T}{N}\]

Then Eq.(\ref{eq:5}) becomes \[R{c_0} = \frac{1}{{{T_0}N{\pi _m}}}\]

For $\pi_m\geq1/N$, we have $Rc_{0}\leq1/T_0$, equivalently
\begin{equation}\label{eq:6}
T_{0}Rc_{0}\leq1
\end{equation}
Eq.(\ref{eq:6}) is the constraint relations between average transmission time and transmission capacity.
\\
Therefore, there is a contradiction between transmission time and transmission capacity. If we want less time, we may not have high transmission capacity. If we want high transmission capacity, we may fail to deliver the packets quickly. This is consistent with the physical intuition.  We should find a routing that can relieve the congestion while considering the transmission time\cite{YanGang,WangWenXu,WangWenXu2,LingXiang,LingXiang2,Bogdan,ZhiHong}.
\\
Eq.(\ref{eq:5}) and Eq.(\ref{eq:6}) seem to suggest : When the stationary distribution of the load of each vertex is near uniform, the corresponding routing, or the corresponding matrix $P$ may have a good transmission capability. A network with the routing of a doubly-stochastic matrix whose stationary distribution ${\pi _i} = 1/N = {\pi _m}$ should have a good transmission capability. The doubly stochastic matrix refers to a transition matrix that the sum of rows and columns are both 1\cite{Inequalities}.while Scale-Free networks are little homogeneous for the degree of each vertex differs a lot.
\\
It should be noted that, although in this paper we study scale-free networks as an example, scale-free is not is a necessary prerequisite for Eq.(\ref{eq:0})(\ref{eq:2})(\ref{eq:3})(\ref{eq:6}).

\begin{figure*}[!ht]\centering
  \includegraphics[scale=0.55]{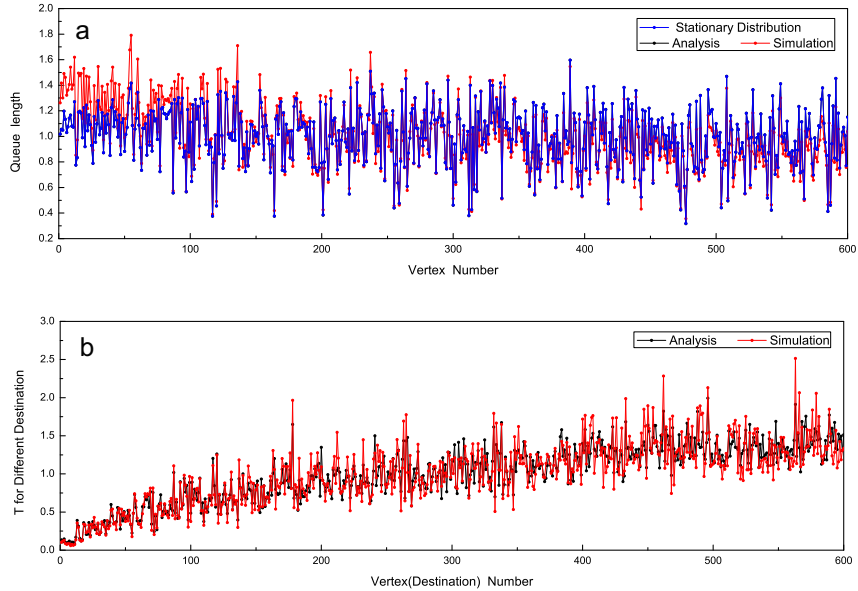}
  \caption{(color online) We select a BA network with $N=600$, $<k>=10$ randomly. FIG.\ref{fg:2}.a shows the average queue lengths of each vertex by analysis and simulation, and the stationary distribution of the transition matrix by SLR.  FIG.\ref{fg:2}.b shows the comparison of the average transmission time of each vertex (destination) between analysis and simulation. In FIG.\ref{fg:2}.a, the graphs of analysis and the stationary distribution of transition matrix are almost overlapping. We normalize the data both in  FIG.\ref{fg:2}.a and b by their own mean value. $C=10,R=35$. Actually, as long as $R$ is remarkably smaller than $Rc$, the graphs in FIG.\ref{fg:2} are irrelevant to $R$.}\label{fg:2}
\end{figure*}
\subsection{\label{2.3}The Estimate of optimum and upper bond }
As mentioned above, there is a contradiction between transmission time and transmission capacity. Only after given one of them we can compare it with the other.
\\
Computer simulation results show that, for a given network, given $T_0$, we can optimize the transmission capacity $Rc_0$~of a local static routing. We denote the maximum $Rc_0$ as $Rc_{0m}$. Obviously, $Rc_{0m}$ is a function of $T_0$. With the variation of $T_0$, it is certain that the maximum of $Rc_{0m}$ exists because the function is bounded.
\begin{figure}[!ht]
\centering
\includegraphics[scale=0.2]{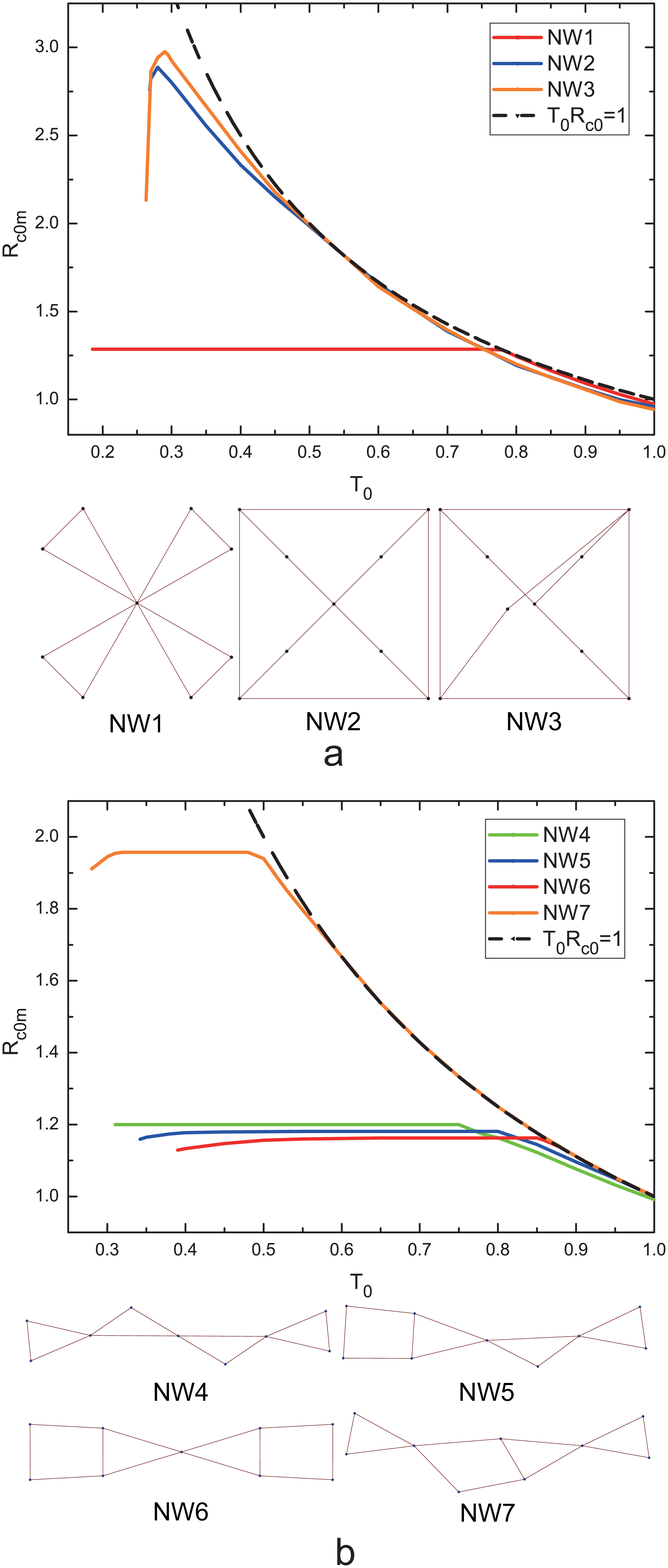}
 \caption{\ref{fg:3}.a (color online) The numbers of vertexes and edges of networks NW $1\sim3$ are equal, but the maximum transmission capacity of these three networks are totally different. The average length of shortest paths of NW1 is the shortest among them three, so $T_0$ can be very small. However, there exists a hub vertex which seriously limited the transmission capacity. Comparing NW3 and NW2, we further reduce the hub property of the centre point of NW2, and improve the transmission capacity a little.\\
 \ref{fg:3}.b (color online) The numbers of vertexes, edges and the average length of shortest paths of networks NW$4\sim7$ are equal, but the graphs of NW7 and NW$4\sim6$ differ a lot. Here we can see: the critical point of the congestion is the vertex in the middle. There is little difference if we change the network structure except the middle vertex. Only when we reduce the hub property of the hub vertex, can we improve the transmission performance greatly.}\label{fg:3}
\end{figure}

Fig.\ref{fg:3} show us: although the routing with best congestion characteristics (max $Rc$) is a routing that each vertex loads uniformly, it is not a completely uniform load routing.
\\
Whether a routing with completely uniform load exists in a network (if and only if $T_0 Rc_0=1$) and what value $T_0$ can be are related to the structure of the network.
\\
Next we estimation the upper bound of $Rc$ of the local routing.
\\
When any $k_i/N$ is an infinitesimal,for sparse networks with very large scale $N$, ($number~of~all~edges/N<some~finite~constant$), there is few differences between $Pd_{i}$ and $P$. So we have the approximation: $Pd_{i}^T\pi\approx C_{i}\pi$, where $\pi$ refers to the stationary distribution of $P$ and $C_i$ refers to a positive constant near $1$.
Considering that we want a routing with large $Rc$, many computer simulation experiments tell us that the loads with the routing are nearly uniformed when $Rc$ is larger, that is :the angle between the all $1$ vector and $\pi$~is lesser.
\\
Basing on the above two conclusions, we approximately have:
\[\frac{{{{(P{d_i}^T)}^n}1}}{N} \approx {({{1^T}P{d_i}^T\pi})^n}\pi\]
With this approximation, we have:\[R{c_0} = \frac{N}{{{\pi _m}\sum\limits_i {(1/\sum\limits_j {{a_{ij}}{\pi _j}} )} }} \le \frac{1}{{ < 1/k > }} \le  < k > \]
Where $<\frac{1}{k}>$ refers to the average of degree¡¯s reciprocals and $<k>$ refers to average degree. For large BA networks, we have $1/<\frac{1}{k}>\approx=3<k>/4$\cite{13,14}. In fact, $\pi _m\sum\limits_i {(1/\sum\limits_j{a_{ij}}{\pi _j} )}$ has a clear physical meaning.
\\
\begin{figure}[!ht]\centering
  \includegraphics[scale=0.3]{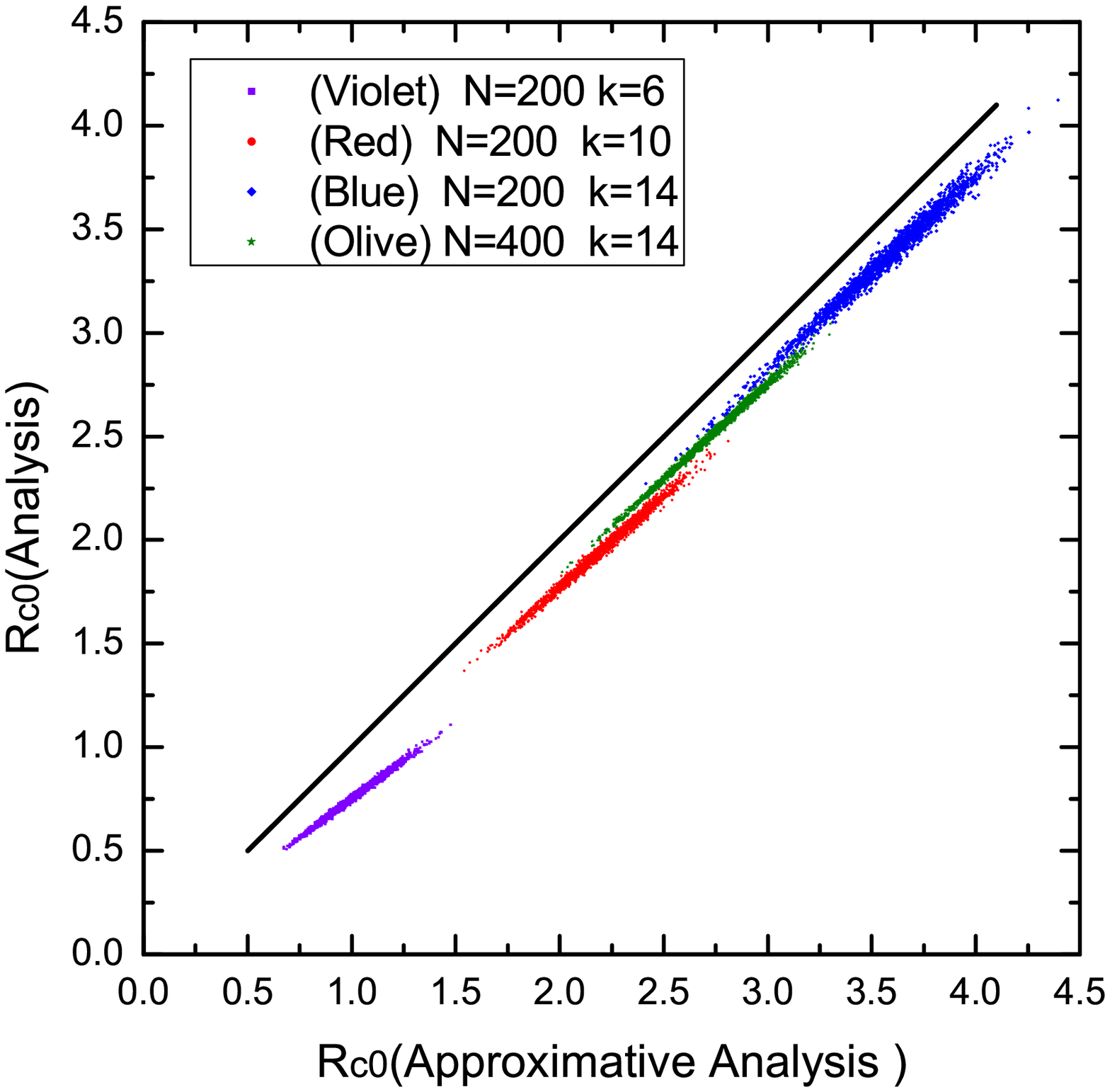}
  \caption{The ordinate is from Eq.(\ref{eq:1}) and the abscissa is the approximate solution of it under the two approximate conditions above. With each parameter, we randomly generate 100 BA networks and 20 transition matrices for each. Thus, we have 2000 data points for each parameter in the graph, the black full line is diagonal, and data points are scattered near it, which supports the validity of approximation. When the network scale is not large but the average degree is not small (e.g. $N=200$, $k=14$), the data points are scattered widely. It also suggests that we can only approximately determine $Rc$ by stationary distribution and network structure for large sparse networks.}\label{fg:4}
\end{figure}
Here can we make an extrapolation of the results? If a vertex knows the information of its neighbours around within $n$ layers ($n$ is not very large), do we have the conclusion~$Rc\leq1/<\frac{1}{k^{(n)}}>$? A vertex's neighbours around within $n$ layers mentioned here is the vertexes that the shortest distance between which and that vertex is $n$. $k^{(n)}$ refers to $n$-layer degree. This extrapolation needs further work to prove.
\section{\label{4}global static routing and betweenness}
The shortest path routing widely used in network communication is a \textquoteleft Link Type\textquoteright~ routing strategy based on global information\cite{YanGang,LingXiang2,Bogdan,ZhiHong}. Simply put, there is a fixed transmission path or \textquoteleft link\textquoteright~ for any packets which are born in vertex $i$ and whose target is vertex $j.$ Each \textquoteleft link\textquoteright~ is the shortest path in some sense. Measuring the link length in different ways, we get the shortest path routing in different meanings.
\\
\cite{YanGang} define these metrics by vertex degree , \cite{LingXiang2} define these metrics by the queue length of each vertex at the moment. The routings in these papers can be considered as shortest routing strategies under different metrics. Of course, these routings all belong to the frame type definition we mention above.
There is a a one-to-one relationship between the language of \textquoteleft link\textquoteright~ and the language of transition matrix for a routing.
\\
Considering a global static routing, now we suppose the routing has been converted into the language of transition matrix. As mentioned above, we have N transition matrices $\{P_1,P_2,P_3\cdots P_N\}$. $P_i$ refers to transition matrix that is used in transferring packets whose target is vertex $i$. Then according to the approximation above, let:
\[{P_i}^{'} = (I - diag({a_{i1}},{a_{i2}} \cdots {a_{iN}})){P_i}\]
then
\[{\beta _{0i}} =(I - {\rm{ }}P_i^{'T})^{ - 1}J_i\]
That is to say, the global routing has $N$ transition matrix rather than $1$, which is different from the local routing. However, $Rc = C/{{{\left| {\left| {{\alpha _0}} \right|} \right|}_\infty }}$, $T = \sum\limits_i {\sum\limits_j {{\alpha _{0ij}}} }$ these two equations are still true. Thus, we present a unified solution for local and global routing.
\\
For the analysis of $Rc$ of  global routing, paper \cite{YanGang} present the results:
\begin{equation}\label{eq:9}
Rc = \frac{{CN\left( {N - 1} \right)}}{{{B_{max}}}}
\end{equation}
where $B_{max}$ refers to the maximum betweenness. Betweenness of a vertex is the number of paths through the vertex.(Strictly, we should consider the probability going through each vertex, but it makes no difference)
Actually, this result matches ours. We can get the betweenness of vertex $i$ ,that is : $Bi=N(N-1)s_{0i}$
by simple calculations.Here $s_{0i}$ refers to the sum of all elements of $i$th row of matrix $\alpha_0$.Thus, we can define betweenness in local routings the same as in global routings.
\\
Now we rewrite the results in local static routings we obtained above using the language of betweenness:
\[\begin{array}{l}
Rc = \frac{{CN(N - 1)}}{{{B_{max}}}},{L_i} = \frac{{R{B_i}}}{{N(N - 1)}}\\\\
T = \frac{{\sum\limits_i {{B_i}} }}{{N(N - 1)}},{T_0} = \frac{B}{{{B_{max}}R{c_0}}}
\end{array}\]

Where $B$ refers to average betweenness.
\\
When $N$ is extremely large, the stationary distribution is approximately proportional to betweenness.
\[{\pi _i} = \frac{{{B_i}}}{{\sum\limits_j {{B_j}} }}\]
Here we can see more clearly, betweenness is consistent with stationary distribution in Markov process, which represent the average number of times that packets go through some vertex per unit time.
\\
It is worth emphasizing that we take the approximation that using \textquoteleft transmission hops\textquoteright instead of \textquoteleft transmission time\textquoteright in the equations expressed in betweenness above. So there is some deviation. However, as the assumptions that satisfies Eq.(\ref{eq:1}) above, this deviation will not be unacceptable most of the time. Significantly, these equations expressed in betweenness are enough to describe the variation of these important variables with different parameters, whatever the accuracy.
\\
It is not difficult for us to derive that in global routings, we also have the equation:\[T_0Rc_0\leq1\]
The equation is the same as that in local routing. When the queue length of each vertex tends to be equal, the inequation becomes an equation.
\\
For local routing, when $T_0$ is small, there nay not exist any routings with which the queue lengths tend to be uniform. For global routings, when $T_0$ is small, there may still exist such routings. This illustrate the reason why transmission capacity of global routings can be much higher than that of local routings from a point of view.
\\
Next, we give an estimate of the upper bound of the divergence speed of global routing $Rc$ with $N$.
From Eq.(\ref{eq:9}) we know $Rc\leq\ CN(N-1)/B$, $B$ is larger than average betweenness $B_0$ in the shortest path routing strategy.
\\
Let the network average shortest path length be $Z$. We have $B_0=(N-1)Z$ (more generally, if packets are not born in each vertex equiprobably, we should calculate it from weighted averages)and
\begin{equation}\label{eq:10}
Rc\leq CN/Z
\end{equation}
That is to say, $Rc$ is limited by network average shortest path length. Similarly, average transmission time $T\geq Z$.
\\
The significance of Eq.(\ref{eq:10}) is that a network with shorter diameter may always have better a transmission performance. This is not an absolute conclusion, because, if there are many vertexes with large degree in a network, they may form some congestion centers and the transmission capability may be limited. If a network does not have small world characteristics, which means $Z\sim N$, with the increasing of $N$, $Rc$ will be limited by a constant. When a network has small word characteristics\cite{17,18,19}, which means $Z\sim N^r$, $0\leq r<1$, with $N$ tending to infinity, $Rc$ tends to infinity. Specially, for BA networks, roughly speaking, we have $Z\sim logN/loglogN$ \cite{13,20,21}. Then divergence speed of $Rc$ will be limited by $NloglogN/logN$ with $N$ increasing to infinity.
Although the above discussion is on static routings, when a dynamic routing (approximately) tends to a stationary state (transition matrix of which tends to be some constant matrix), our method is still suitable. The reason why dynamic routings have better performance than static routings is that dynamic routings have better adjustability (more parameters and more information). Besides helping us to find a routing with better performance, the better adjustability leads to the better adaptability to transmission conditions. Thus dynamic routings are more practical. Tending to a stationary state is often a basic requirement for a dynamic routing, for people need stable transmission. Thus, the significance of our paper is not only limited in static routings.
\section{\label{5}conclusion}
The problem of data transmission on network is essentially a problem of a queuing system. We are not the first ones to use Markov processes in modeling of queuing systems. It is a successful application of Markov processes in this paper and we get some new and basic results.
For all local and global static routings, we get a universal inequality $Rc_0T_0\leq1$. For all local static routings, with certain assumption, we have $Rc\leq1/<1/k>$.

\bibliography{s2}

\end{document}